\documentclass[11pt, draftclsnofoot, onecolumn]{IEEEtran}

\usepackage{cite,graphicx,amsmath,amssymb}
\usepackage{subfigure}
\usepackage{citesort}
\usepackage{fancyhdr}
\usepackage[table]{xcolor}
\usepackage{array}
\usepackage{mdwmath}
\usepackage{mdwtab}
\usepackage{balance}
\usepackage{xcolor}
\usepackage{bm}
\usepackage{amsthm}
\usepackage{amssymb}
\usepackage{threeparttable}
\usepackage{algorithm}
\usepackage{algorithmic}
\usepackage{multirow}
\usepackage{flafter}

\begin{document}
\title{Artificial Intelligence Aided Next-Generation Networks Relying on UAVs}
\author{
Xiao~Liu,~\IEEEmembership{Student Member,~IEEE,}
Mingzhe~Chen,~\IEEEmembership{Member,~IEEE,}\\
 Yuanwei~Liu,~\IEEEmembership{Senior Member,~IEEE,}
        Yue~Chen,~\IEEEmembership{Senior Member,~IEEE,}\\
        Shuguang Cui,~\IEEEmembership{Fellow,~IEEE,}
        and Lajos Hanzo,~\IEEEmembership{Fellow,~IEEE,}



}

\maketitle

\begin{abstract}

Artificial intelligence (AI) assisted unmanned aerial vehicle (UAV) aided next-generation networking is proposed for dynamic environments. In the AI-enabled UAV-aided wireless networks (UAWN), multiple UAVs are employed as aerial base stations, which are capable of rapidly adapting to the dynamic environment by collecting information about the users' position and tele-traffic demands, learning from the environment and acting upon the feedback received from the users. Moreover, AI enables the interaction amongst a swarm of UAVs for cooperative optimization of the system. As a benefit of the AI framework, several challenges of conventional UAWN may be circumvented, leading to enhanced network performance, improved reliability and agile adaptivity. As a further benefit, dynamic trajectory design and resource allocation are demonstrated. Finally, potential research challenges and opportunities are discussed.


\end{abstract}


\section{Introduction}

Owing to their agility, as well as their ability to establish line-of-sight (LoS) wireless links, \textit {unmanned aerial vehicles (UAVs)} have become a focal point in the wireless communications research field for mitigating a wide range of challenges encountered in diverse commercial applications~\cite{Wang2017taking}. Given these beneficial characteristics of UAVs, they can be used as aerial base stations (BSs) to complement and/or support existing terrestrial communication infrastructure since they can be flexibly redeployed in temporary traffic hotspots (e.g., political rally, concert hall, sports stadium) or after natural disasters~\cite{zeng2016wireless}. Thus, UAV-assisted wireless networks (UAWN) have been successfully applied in UAV-assisted emergency communications (UAV-EC), UAV-aided cellular offloading (UAV-CO), and UAV-assisted Internet-of-Things (UAV-IoT) systems.


To effectively exploit UAVs for assisting wireless networks, early research contributions have studied a number of technical challenges~\cite{Fotouhi2019Survey} that include three-dimensional (3D) deployment, trajectory design, interference management and resource allocation. Powerful optimization techniques, such as convex optimization~\cite{Wang2018Joint}, game theory~\cite{yan2018game}, transport theory~\cite{mozaffari2017wireless} and stochastic optimization~\cite{liu2019uavm} have been invoked for addressing the aforementioned fundamental challenges. In the conventional UAWN, the links between the UAVs and the users are typically modeled as dominant LoS channels. This assumption converts many of the associated problems to convex problems, which are readily solved by conventional convex optimization algorithms. However, the dominant LoS channels are often blocked by high-rise buildings in the practical application of UAVs, especially in city-center scenarios. Moreover, the users in conventional UAWNs are often assumed to be static for simplicity, i.e. the mobility of users is typically ignored. Another limitation in the existing literature is that the UAVs are not capable of learning from the environment or from the feedback of the users for further enhancing the service quality. In practical applications of UAVs in wireless networks, the system parameters are treated as random variables, which naturally leads itself to the derivation of insightful joint probability distributions conditioned on the users' tele-traffic demand and mobility. However, this is a high dynamic stochastic environment, which constitutes quite a challenge for conventional optimization approaches. Finally, simultaneously employing a swarm of UAVs becomes more challenging due to the cooperation amongst UAVs and owing to the inter-cluster interference, which aggravates the challenge imposed on conventional optimization.

Given these challenges, artificial intelligence (AI) aided optimization comes to rescue\cite{yao2019AI}. More explicitly, big data analytics and machine learning (ML) may be invoked for tackling the high-dynamic design problem of UAWNs. Table I provides a summary of the challenges and application scenarios of UAWNs, additionally highlighting, where ML-based solutions may become beneficial. By exploiting the learning capability of ML, the aforementioned challenges encountered in UAWNs may be mitigated, leading to improved network performance. Against this background, in this article, we highlight the key features of AI-enabled UAV networks, as well as how AI improves the performance of UAWNs. The new contributions of this paper compared to the state-of-the-art is provided in Table II.

\begin{table*}[htbp]\footnotesize
  \begin{center}
\begin{threeparttable}
\caption{Challenges, applications, and ML-based solutions for UAWNs}
\centering
\begin{tabular}{|l||c||c||c||c||c||c||c||c||c|}
\hline
\centering
\multirow{2}{*}{Challenges in the UAWN} &
\multicolumn{3}{c|}{Application scenarios} &
\multicolumn{6}{c|}{ML-based solutions}  \\
\cline{2-10}
  & UAV-EC & UAV-CO & UAV-IoT &  QL & DL & DQN & DDPG & EA & SARSA \\
\hline
\centering
3D deployment & \checkmark & \checkmark & \checkmark & \checkmark & \checkmark & \checkmark &  &  & \checkmark \\
\hline
Trajectory/path planning & \checkmark & \checkmark & \checkmark & \checkmark & \checkmark & \checkmark & \checkmark & \checkmark & \checkmark \\
\hline
Resource allocation & \checkmark & \checkmark & \checkmark & \checkmark & \checkmark & \checkmark &  & \checkmark & \\
\hline
Interference management & \checkmark & \checkmark & \checkmark & \checkmark & \checkmark &   &   &  & \\
\hline
3D channel model & \checkmark & \checkmark & \checkmark &   & \checkmark  &   &   &  & \\
\hline
Energy management & \checkmark & \checkmark & \checkmark & \checkmark &   &  \checkmark &   &  & \\
\hline
Feature extraction & \checkmark & \checkmark &  &  & \checkmark &   &   &  & \\
\hline
Security & \checkmark & \checkmark &  & \checkmark & \checkmark &   &   &  & \\
\hline
\end{tabular}
\begin{tablenotes}
\item[1] QL represents Q-learning; DL is short for Deep Learning; DQN represents a Deep Q-Network; DDPG is the acronym for deep deterministic policy gradient base algorithms; EA represents evolutionary algorithms; SARSA represents state-action-reward-state-action algorithms.
\end{tablenotes}
\end{threeparttable}
\label{application}
  \end{center}
\end{table*}

\begin{table*}[htbp]\footnotesize
  \begin{center}
\begin{threeparttable}
\caption{New contributions of this paper compared to the state-of-the-art}
\centering
\begin{tabular}{|l||c||c||c||c||c||c||c|}
\hline
\centering
  & this paper& [1]-2017 & [2]-2016 & [3]-2019 & [10]-2018 & [11]-2019 & [12]-2019  \\
\hline
AI architecture & \checkmark &   &   &   &   &   &   \\
\hline
Big data methods & \checkmark &   &   &   & \checkmark & \checkmark &    \\
\hline
ML-based solutions & \checkmark & \checkmark &   &   & \checkmark & \checkmark & \checkmark  \\
\hline

\end{tabular}
\end{threeparttable}
\label{contributions}
  \end{center}
\end{table*}

\section{AI-Assisted UAV-Aided Networking}

\begin{figure*}
\centering
\includegraphics[width=6in]{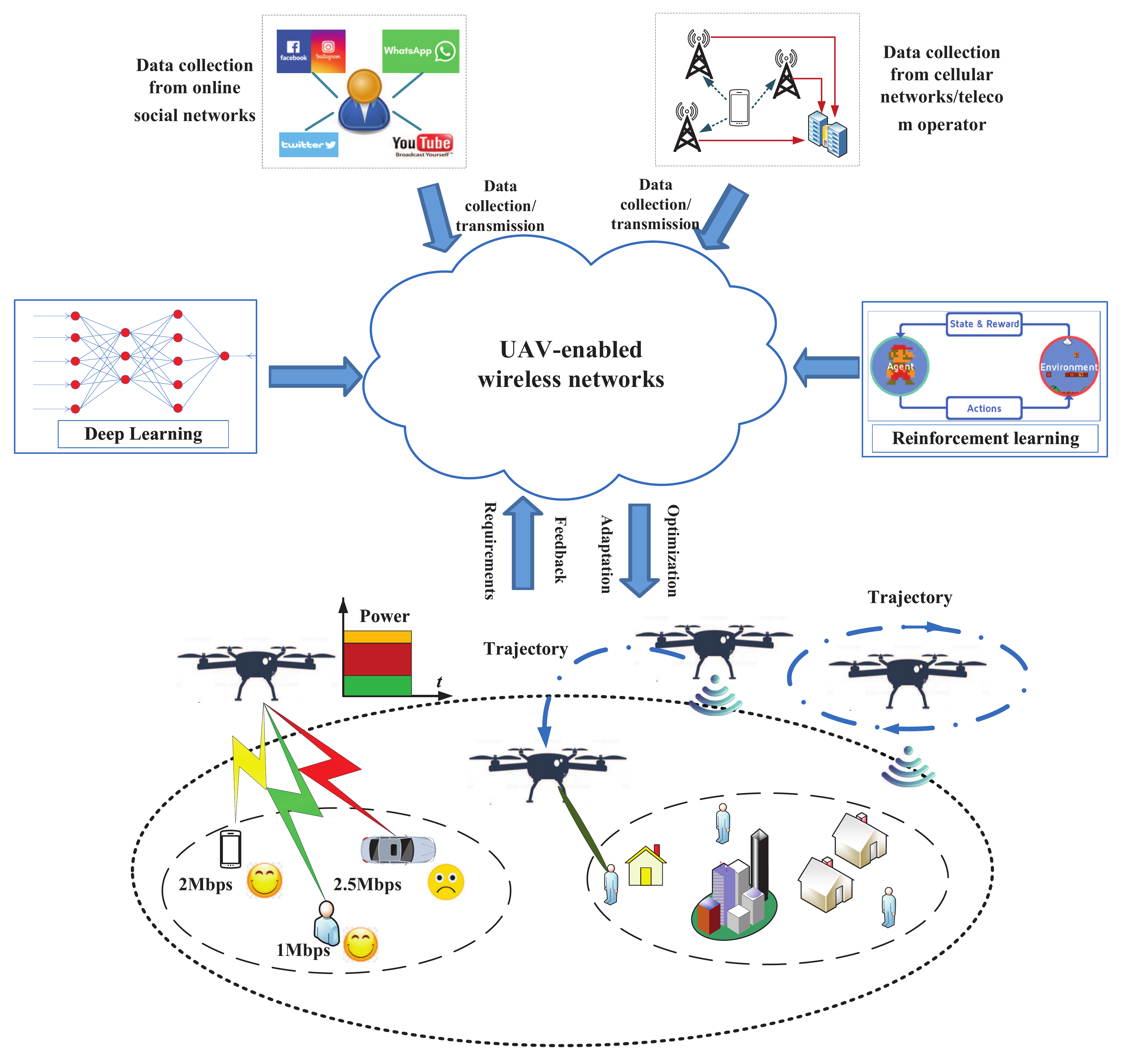}
\caption{Architecture of AI-enabled UAV networks.}\label{Architecture}
\end{figure*}

In this section, we first present the system architecture of our proposed AI-enabled UAWN, followed by its key modules.

\subsection{Key Features for AI-enabled UAWN}

Conventional UAWNs are designed based on an analytical model that can be optimized in an iterative manner. However, wireless networks operate in a complex time-variant environment, where the classic mathematical models have limited accuracy, which may potentially be improved by sophisticated AI techniques.

Fig. 1 illustrates the proposed AI-aided UAWN architecture advocated, where the downlink of the UAWN is considered. Multi-UAV are employed as aerial base stations for supporting the users in a particular area in which the existing terrestrial networks may be limited in capacity or coverage. In our AI-aided UAWN, feature extraction is invoked before optimizing the design of the network. First, the associated user information (e.g., position, data demand, mobility) is collected, stored and processed. Thus, the users' behavior and requirements can be predicted for efficiently controlling and operating the UAVs. Given the feature extracted, adaptive schemes are invoked for network planning, resource allocation, and interference coordination. In this system, the UAVs are capable of rapidly adapting to the dynamic environment by learning both from the environment and from the feedback of the users. Moreover, the cooperation of a swarm of UAVs may be invoked. Among all the challenges encountered by the UAWNs, the geographic UAV deployment/trajectory design and resource allocation problems are perhaps the most fundamental ones, which will be considered below.

\subsubsection{AI-aided dynamic deployment and trajectory design of multi-UAV}

In contrast to the conventional UAWNs, in our AI-aided UAWN, both LoS and non-line-of-sight (NLoS) conditions are encountered, instead of assuming dominant LoS channels for the associated air-to-ground communications. In this context we often arrive at non-convex optimization problems. Sophisticated AI techniques may be invoked for solving these challenging non-convex problems, especially when the mobile users are roaming at speed. Thus, the UAVs have to be periodically repositioned for efficiently serving them. More particularly, in the first step, the users' position information may be acquired with the aid of a real dataset, which consists of the users' GPS-related position information. Subsequently, the dynamic 3D trajectory of the UAVs is designed based on the users' mobility information, while maintaining high service quality and reducing the response time.

\subsubsection{AI-enabled resource allocation for multi-UAV networks}

In contrast to the conventional UAWNs, the specific time-varying tele-traffic requirement of each user may be readily accommodated by AI-aided UAWN. Thus, the required wireless resource (e.g., bandwidth, transmit power, and computational resource) also varies, which emphasizes the importance of agile resource allocation to be carried out by the UAVs for serving the demands of the associated users. In the AI-enabled UAWNs, the users' tele-traffic demand is firstly predicted based on a real dataset, which consists of census information, cellular infrastructure deployment, and cellular data demand. Given the predicted mobile tele-traffic, both unnecessary delays and resource wastage may be avoided. Moreover, due to the limited-capacity battery of UAVs, UAVs must consider energy efficiency while optimizing resource allocation. However, energy efficiency and resource allocation of each UAV are dependent since energy efficiency depends on the resource allocation such as transmit power and computational resource allocation. In the AI-enabled UAWNs, UAVs can find the relationship between energy efficiency, resource allocation, and quality-of-service, and, hence, jointly optimizing energy efficiency and resource allocation.

\subsection{Big Data in Predictive Employment of Multiple UAVs}

Given the explosive proliferation of wireless data services, a rich set of cellular-related data became available in online social networks, as well as from the telecom operators' platforms for acquisition. The emergence of big data analytic techniques also makes it possible to extract and predict the cellular-related information from a significant volume of data that are collected from multiple sources, such as online social networks, cellular networks/telecom operators and from the roadside units infrastructures.

\subsubsection{User mobility information}

To fully exploit the potential gains for enhancing the wireless quality-of-service and for reducing the required communications resources by predicting the users' mobility, online social networks over smartphones, which has accumulated a substantial amount of geographical data, may be relied upon for acquiring the users' mobility information~\cite{liu2019trajectory}. For instance, many social networking applications (e.g., Instagram, Twitter, Facebook) are authorized to collect and share users' locations or GPS coordinates. Thus, the geographic user distribution may be potentially estimated and predicted.


\subsubsection{User data requirement information}

On a similar note, we can rely on the data collected from online mobile applications and on data collected by cellular networks/telecom operators for acquiring the users' data demand. On the one hand, mobile applications (e.g., YouTube, Youku) have been granted privileges for recording information, such as the users' interests and historical requests, which characterize the tele-traffic distribution. On the other hand, the individual users' tele-traffic demand is recorded by the cellular infrastructure or by the mobile internet big data platform of telecom operators~\cite{gu2018multiple}. These big data storage and analysis platforms provide compelling opportunities for enhancing the efficiency of on-demand cellular services.

Apart from the aforementioned features, the results of channel sounder design and channel measurement campaigns can also be leveraged for improving the performance of the networks. The performance of the UAWNs may be further optimized by invoking 3D radio environment maps (including building reconstruction) based on the geographical information system. Hence, the performance of the 3D collaborative UAWN design can be tested in a life-like urban environment for verifying the above claims.

\subsection{Machine Learning methodologies for the UAWN}

Machine learning is one of the key AI-aided research areas. The core idea of the ML-assisted techniques adopted in the UAWN is that they allow the UAVs to improve their service quality by learning from the environment, from their historical experience and from the feedback of the users. Since the collected data are of multi-source, heterogeneous and voluminous nature~\cite{gu2018multiple}, deep learning (DL), which is capable of accurately tracking the state of a network and of predicting its future evolution~\cite{challita2019machine}, can be a promising technique for transforming the data to actionable knowledge. Therefore, it is firstly invoked for predicting both the tele-traffic demand and the mobility of the users. On a similar note, reinforcement learning (RL) has also witnessed increasing applications in the fifth-generation (5G) wireless systems. More explicitly, RL models can be used for supporting the UAVs (agents) in their interaction with the environment (states) and by learning from their mistakes, whilst finding the optimal behavior (actions) of the UAVs. Furthermore, the RL model can incorporate farsighted system evolution (long-term benefits) instead of focusing on current states. Thus, it is invoked for solving challenging problems in UAWNs~\cite{mozaffari2019tutorial}.

\section{AI-Enabled Deployment and Movement Design for Multi-UAV Networks}

To validate the distinguished capabilities of AI in UAWNs, an AI-enabled multi-UAV positioning and trajectory design is considered in this section.

\subsection{Motivations}

Since most of the existing research contributions on positioning and trajectory design of UAVs assume that the users are static, the issues of user-mobility have been neglected. Moreover, typically a dominant LoS connection is assumed between the UAVs and users. These two assumptions convert the statistic dynamic problem to a static optimization problem or even to a convex problem, which falls into the field of conventional optimization algorithms. However, the desirable agility of UAVs is at odds with these unrealistic assumptions. Additionally, most of the conventional trajectory design methods can only function in the idealized scenario, when perfect knowledge is available about both the environment and users. However, they are not capable of learning from the environment or learning the users' behavior. To address the aforementioned challenges, we can resort to invoking AI techniques for the repositioning of UAVs.

\begin{figure}
\centering
\includegraphics[width=85mm]{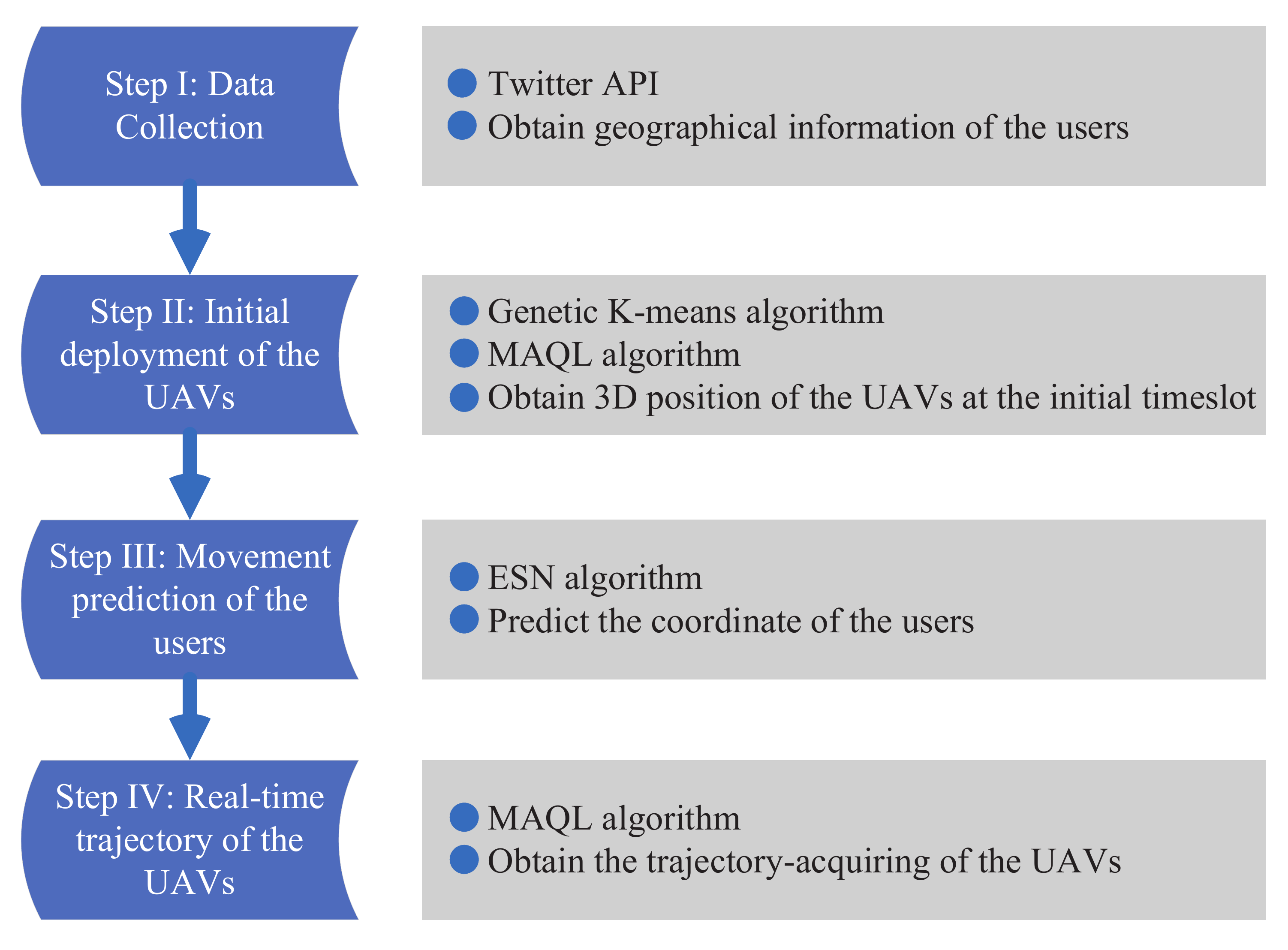}
\caption{The procedure and algorithms invoked for the dynamic trajectory design of multi-UAV.}\label{trajectoryprocedure}
\end{figure}

\subsection{AI-Enabled Deployment and Movement Design}

Based on the system architecture of AI-enabled UAWNs discussed in Section II, let us now consider both the positioning and trajectory design of multi-UAV in this section. The algorithms invoked are illustrated in Fig.2. As mentioned before, the UAVs have to be periodically repositioned based on the users' mobility. To accumulate real mobility information, the relevant coordinate data can be collected from the Twitter API. When Twitter users tweet, the Twitter API is authorized to record their GPS-related coordinate information and make it available to the general public. Thus, the users' movement can be predicted by mining data from the Twitter API. Since the UAWNs rely on a dynamic time-variant model, which is a challenge for conventional optimization solutions. RL and DL algorithms come to rescue, as a benefit of their learning capability for solving the associated dynamic trajectory design problems.

DL is capable of prompting accurate processing, hence we have invoked it for predicting the users' mobility. Taking the echo state network (ESN) algorithm for example, the input of the ESN model is users' position vector collected from Twitter while its output vector is the predicted users' position information. The ESN model aims for training a model with the aid of its input and output to minimize the mean square error with lower complexity. Given the mobility of users, the trajectory of the UAVs can be determined by optimizing the long-term benefits. Again since it is non-trivial to formulate the UAWN problem as a supervised learning problem due to its strong interactions with the environment, the RL algorithm is invoked. For the cooperative deployment of a swarm of UAVs, which is naturally a multi-agent model. Therefore, the multi-agent RL algorithm is chosen for tackling this problem. Taking the multi-agent Q-learning (MAQL) algorithm for example, the states are consisted of three parts: the current 3D position of each UAV; the current 2D position of each user; as well as the current transmit power level of each UAV. At each timeslot, each agent (UAV) carries out an action (choosing a specific direction and transmit power level), after which, the UAV receives a reward/penalty based on the instantaneous sum rate of the users. The global Q-value is decomposed into a linear combination of local agent-dependent Q-values. Thus, if each agent maximizes its Q-value, the global Q-value will be maximized. Finally, by reconstructing 3D radio environment maps of a particular area, the performance of the 3D collaborative UAWN design can be tested in a life-like urban environment for verifying the performance of the proposed solutions.

\begin{figure} \centering
\subfigure[Trajectory design on Google Map.] { \label{fig:a}
\includegraphics[width=0.48\columnwidth]{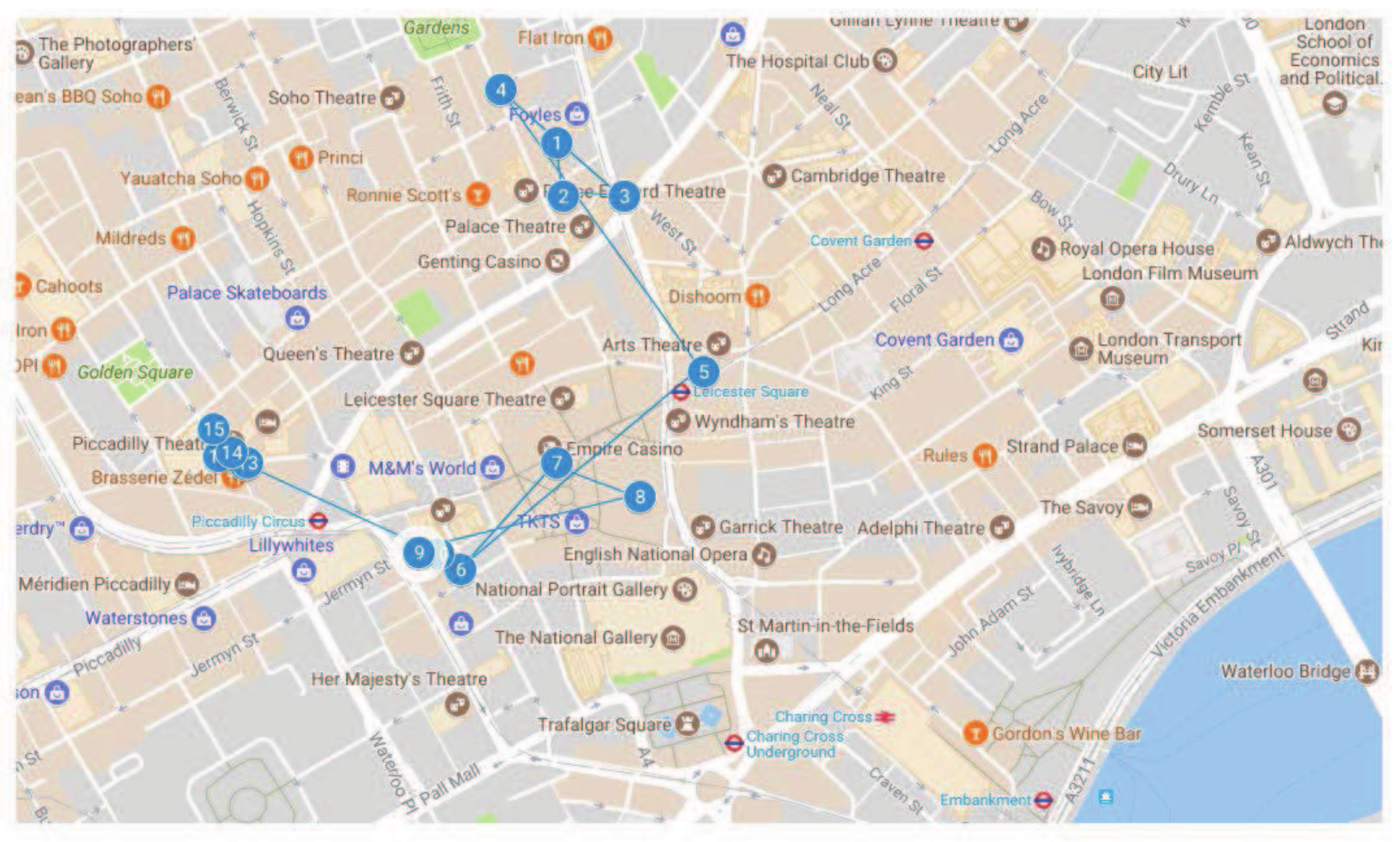}
}
\subfigure[Performance comparison over throughput.] { \label{fig:b}
\includegraphics[width=0.45\columnwidth]{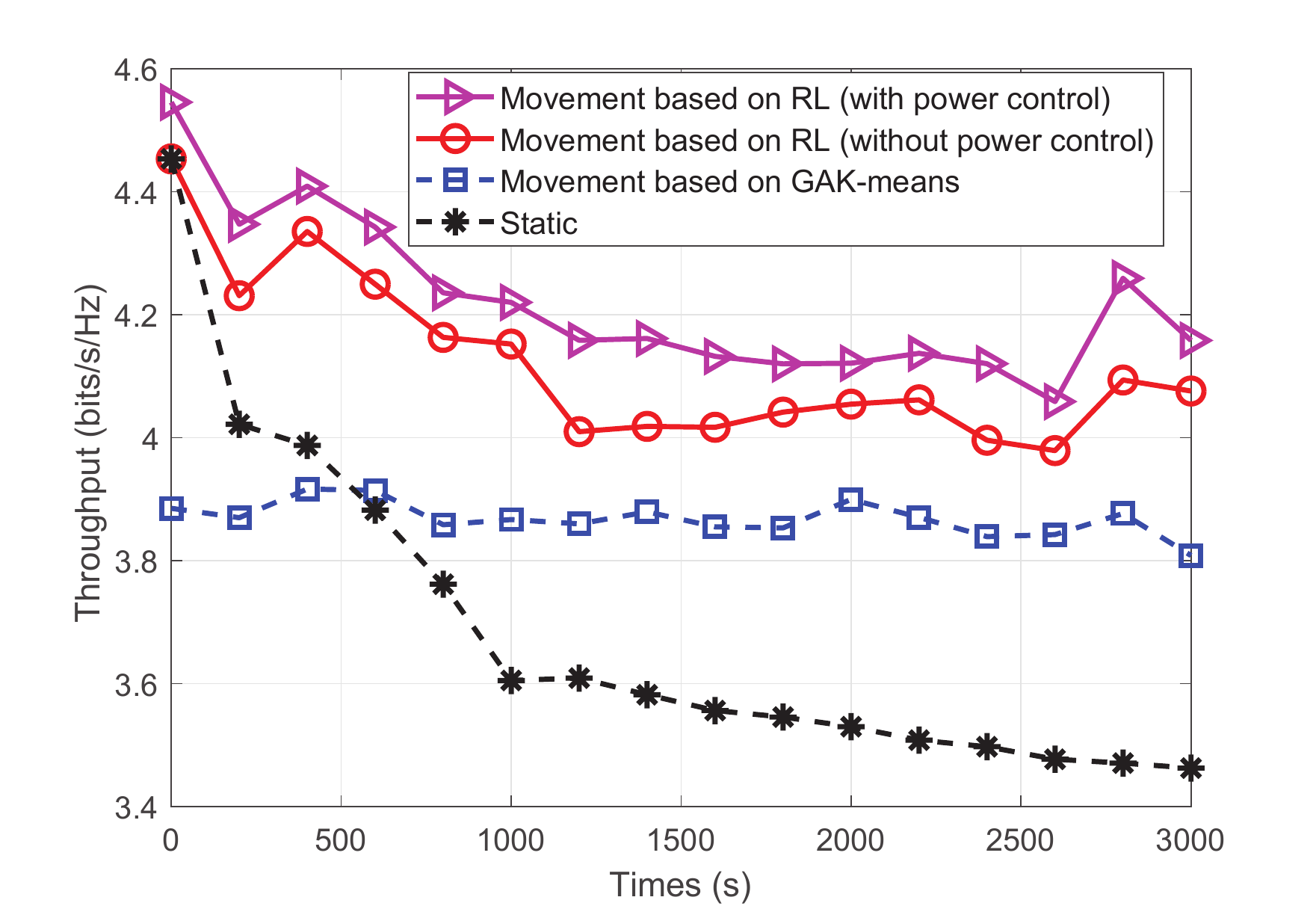}
}
\caption{Performance of the system derived from the centralized MAQL algorithm~\cite{liu2019trajectory}. }\label{esnplot}
\end{figure}

Fig. 3 of the simulation results in~\cite{liu2019trajectory} characterizes the throughput of the UAWN with traveling UAVs, while the trajectory of the UAVs is designed on the Google map based on our Twitter-based real dataset. It can be observed from Fig. 3 that moving UAVs are capable of improving the service quality provided by static UAVs. Additionally, this figure also illustrates that with the aid of accurate transmit power control of the UAVs a high-quality service can be maintained compared to its counterpart dispensing with power control.

\section{AI-Enabled Resource Allocation for UAV networks}

Let us now continue by highlighting the motivation of using AI techniques for resource allocation in UAV-based wireless networks, followed by a use case scenario.

\subsection{Motivations}

Recently, most of the existing contributions that optimize resource allocation for UAV-based wireless networks assume that the transmission links between the UAVs and users are static LoS channels. Hence, the loss of the air-to-ground channel depends only on the distance between the UAVs and users, which may not capture the unique feature of UAVs, such as the altitudes of UAVs and the rainy or snowy conditions that substantially affect the loss of transmission links. Moreover, due to the lack of accurate UAV channel models for the visible light communication and millimeter wave (mmWave) band, most of the existing treatises are focused on resource allocation for UAVs in the sub-6 GHz band. Additionally, most of the conventional resource allocation methods ignored the user behaviors and the wireless environment that may significantly affect all aspects of resource allocation. For instance, by generating the radio map related to the wireless environment, UAVs can optimize resource allocation depending on the potential frequency-domain interference. To address these challenges, we can invoke AI-based optimization techniques for UAVs. In particular, AI techniques enable UAVs to analyze their collected data so as to predict both the wireless environment and the user states. Based on the prediction and analysis results, AI-enabled UAVs can automatically optimize their resource allocation. For example, AI-enabled UAVs can use RL to adjust their resource allocation decisions,  locations, path planning, and flying directions,  according to the users' movements and the change of wireless environment so as to service their ground users optimally.

\begin{figure}[!t]
  \begin{center}
   \vspace{0cm}
    \includegraphics[width=8cm]{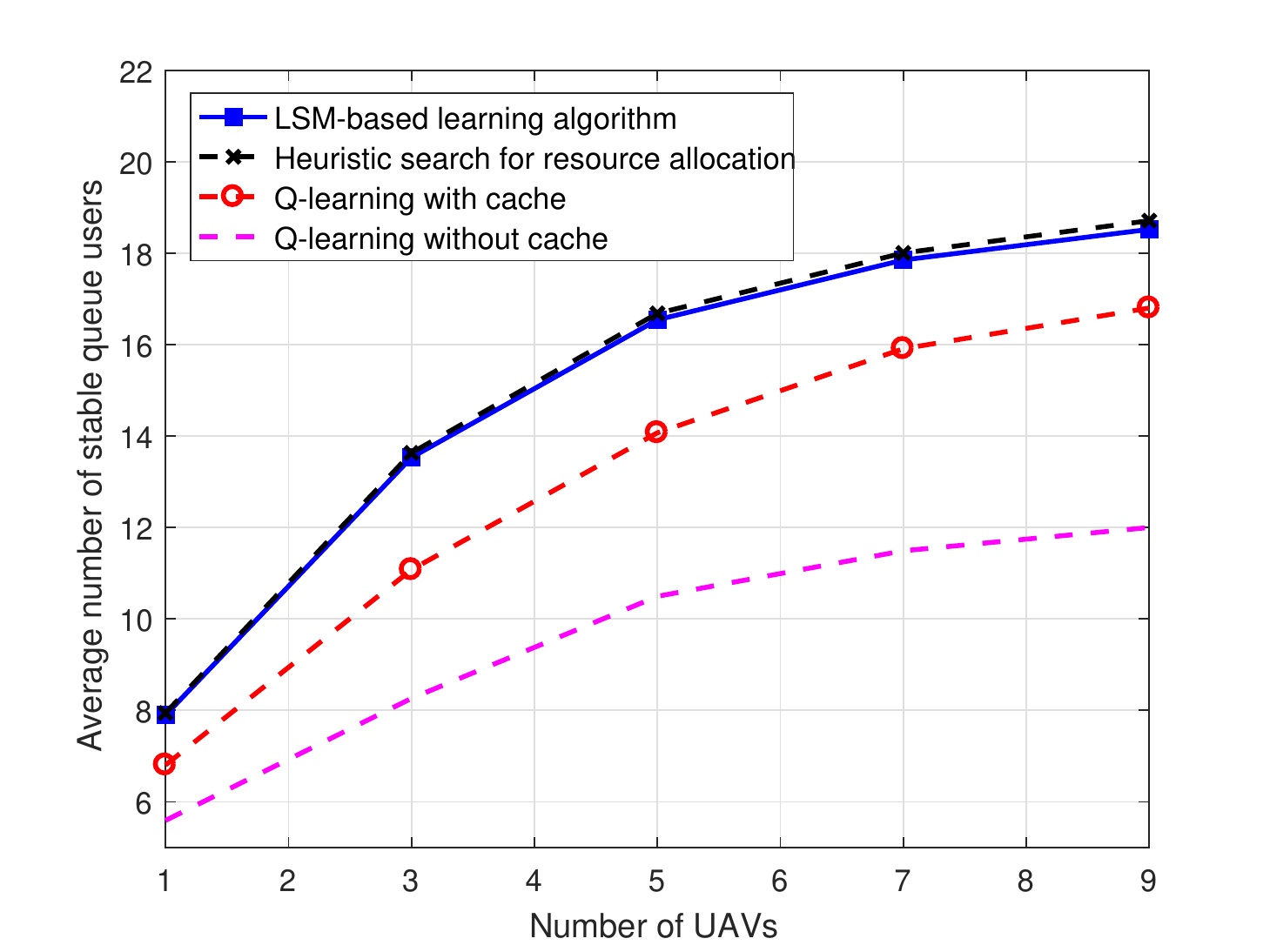}
    \vspace{-0.3cm}
 {\caption{\label{Fig. 1} Average number of stable queue users as the number of AI-enabled UAVs varies\cite{mingzhe2019TWC}.}}
  \end{center}\vspace{-0.8cm}
\end{figure}

\subsection{AI-enabled Resource Allocation}



An elegant appliaction of AI techniques for spectrum allocation within AI-enabled UAV networks is explained in \cite{mingzhe2019TWC}. In \cite{mingzhe2019TWC}, the joint caching management and resource allocation problem is studied for a cache-enabled UAV network in which the UAVs can service users using both the LTE unlicensed (LTE-U) and licensed bands. The contents that the users request can be transmitted from either the cache units at the UAVs directly or via content servers. Since the users' content requests dynamically change, the UAVs must develop an intelligent algorithm to adapt their resource allocation and caching schemes so as to optimally service users. In \cite{mingzhe2019TWC}, A liquid state machine (LSM)-based RL algorithm is proposed for optimizing spectrum resource over Sub-6 GHz and LTE over unlicensed (LTE-U) bands as well as finding optimal contents to cache. Using the LSM-based approach, AI-enabled UAVs can find the optimal policies for the optimal contents to store at UAV cache, spectrum allocation, and user association, when the wireless environment states and the users¡¯ content requests change dynamically. This is because an LSM can build the relationship between the content requested by each user and the content caching, user association, resource allocation schemes, as a benefit of its large memory.


Fig. \ref{Fig. 1} of the simulation result of \cite{mingzhe2019TWC} shows how the average number of users that have stable queues varies, as the number of AI-enabled UAVs changes. From Fig. \ref{Fig. 1}, we show that, as the number of AI-enabled UAVs increases, the number of users that have stable queues increases. This stems from the fact that, as the number of UAVs increases, the users have more connection options, and, hence, improving the number of users that have stable queues. In Fig. \ref{Fig. 1}, we can also see that, compared to the Q-learning having a cache and to Q-learning operating without cache, the LSM-based learning approach can improve {17.8\%} and {57.1\%} gains of the number of users that have stable queues for a network supported by 5 UAVs. This is due to the fact that the LSM based learning approach can exploit the historical information of the ground users to predict the users' content request distribution so as to optimize caching contents and to find an optimal user association scheme.

\section{Concluding Remarks and Future Challenges}

\subsection{Concluding Remarks}

In this paper, we have analyzed the prospects of AI techniques in UAV-assisted wireless networks. Both big data aided feature extraction and ML-aided optimization solutions are employed for enhancing the service quality of the users. The key benefits of AI-enabled networks were identified compared to conventional UAWN. This was followed by introducing a pair of UAWN case studies, namely the positioning and dynamic trajectory design as well as the dynamic resource allocation of multi-UAV networks.

\subsection{Challenges of Using AI for UAV-based Wireless Networks}

\subsubsection{Distributed solutions for the AI-enabled UAWNs}

Since the cooperative deployment/trajectory design and resource allocation of a swarm of UAVs was considered, intensive communication and coordination among the UAVs or between the UAVs and the ground control center are required in the centralized MAQL model. Each UAV has to maintain a Q-table that includes data both about its states as well as about the other UAVs' states and actions. Naturally, additional communication resource is required for coordination among the UAVs. To alleviate this limitation, decentralized approaches that are capable of making decisions and taking actions in a distributed manner can be invoked for maximizing the long-term benefits. Therefore, each UAV decides the optimal position or allocated resource of itself without any information exchange with other UAVs.

\subsubsection{Uninterruptible wireless supply for the AI-enabled UAWN}

Battery-powered as it is, the energy constraint of the UAVs is another core challenge in the AI-enabled UAWN, especially when on-board data processing and on-line computation are used. Moreover, bad weather such as strong breeze, excessively cold or hot conditions can also drain the batteries more quickly. Since increasing the battery size is impracticable due to the size and weight constraints, frequent battery recharge is expected. To alleviate this limitation, energy-efficient designs have to be conceived for AI-enabled UAWNs by considering the propulsion energy. An alternative is to recharge the UAV by wireless power transfer or laser charging using laser-guns at the roof-tops~\cite{liu2016Charging}. However, uninterruptible wireless supply form UAVs has not as yet been realized.

\subsection{Future Works}

\subsubsection{AI in UAV-assisted wireless networks with NOMA}

The key idea of power-domain non-orthogonal multiple access (NOMA) is to superimpose the signals of two users at different powers for exploiting the spectrum more efficiently by opportunistically exploring the users' different channel conditions. By invoking NOMA for pursuing further throughput enhancement, and massive wireless connectivity, the UAWN scenario becomes an attractive multi-cell downlink NOMA transmission model. Feature extraction can be invoked for predicting the tele-traffic demand of users, which aims for identifying potential tele-traffic congestion events, while the multi-agent RL algorithm can be leveraged for determining the trajectory and power allocation of the UAVs, as well as the UAV-user association.

\subsubsection{AI in UAV-assisted vehicular networks}

The UAVs can be invoked for assisting the vehicular networks by forming cooperative air-to-ground vehicular networks. The vehicular subnetwork on the ground is enhanced with the aid of the aerial subnetwork formed by the multi-UAV layer. In cooperative vehicular networks, UAVs are not only employed as aerial BSs but are also capable of collecting traffic information from areas that are inaccessible for ground vehicles or roadside units due to bad weather or lighting conditions. Moreover, the UAVs are also capable of acting as intermediate relays for enhancing the connection among vehicles as well as between vehicles and the infrastructures.

\subsubsection{AI in charging solutions for UAVs}

By mounting a compact distributed laser charging (DLC) receiver or wireless power transmission (WPT) receiver antenna on UAVs, complemented by a DLC/WPT transmitter (termed as a power base station) on the ground, building roof, or even on mobile vehicles, the UAVs can be charged as long as they are flying within the coverage range of the DLC/WPT transmitter. Thus, these DLC/WPT-aided UAVs can operate for a long time without landing until maintenance is needed. The dynamic charging problem can be modeled as a Markov decision process (MDP), where the optimal number and position of the charging stations, the trajectory of both the UAVs and the MPC, as well as the UAV-station association can be optimized with the aid of AI solutions.

\subsubsection{AI in multi-tier and multi-functional UAV networks}

In the future wireless networks, the UAVs can also act both as content providers and computing servers. Diverse categories of UAVs and different sizes of UAVs are deployed at different altitudes for disparate missions. Thus, multi-tier and multi-functional UAV networks can be formed. The optimization for resource allocation (including radio, cache, energy and computing resources), trajectory and user association has to be achieved jointly in a dynamic environment. Collective artificial intelligence relies on multi-category AI agents for achieving the same goal based on local training at each AI agent, with limited or no direct communication amongst the agents. Thus, the UAVs can make a concerted effort to improve the performance of UAWNs.

\subsubsection{AI in intelligent reflecting surface assisted UAV networks}

Intelligent reflecting surfaces (IRS) are capable of proactively 'reconfiguring' the wireless propagation environment by compensating the path-loss over long distances, as well as for forming virtual LoS links between the UAVs and users via passively reflecting their received signal. Due to the intelligent deployment and design of the IRS, a software-defined wireless environment may be constructed, which in turn, provides potential received SINR enhancements. The throughput enhancement attained becomes more considerable when the LoS link between the UAVs and users is blocked by high-rise buildings. Thus, the performance of UAWNs may be further improved.

\bibliographystyle{IEEEtran}

 \end{document}